\documentclass[11pt]{article}

\usepackage[margin=1.1in]{geometry}
\usepackage{amsmath,amssymb,amsthm}
\usepackage{bm}
\usepackage{booktabs}
\usepackage{graphicx}
\usepackage{caption}
\usepackage{subcaption}
\usepackage[round]{natbib}
\usepackage{setspace}
\usepackage{xcolor}
\definecolor{linkblue}{RGB}{25,60,120}
\usepackage[colorlinks=true, linkcolor=linkblue, citecolor=linkblue,
            urlcolor=linkblue]{hyperref}
\usepackage{enumitem}
\usepackage{microtype}

\graphicspath{{figures/}}

\newtheorem{hypothesis}{Hypothesis}
\newtheorem{proposition}{Proposition}

\newcommand{\E}{\mathbb{E}}
\newcommand{\Prob}{\mathbb{P}}
\newcommand{\R}{\mathbb{R}}
\newcommand{\F}{\mathcal{F}}
\newcommand{\lin}{\lambda_{\mathrm{in}}}
\newcommand{\lout}{\lambda_{\mathrm{out}}}

\title{\textbf{LLM-Enhanced Dynamic Financial Knowledge Graphs for
Cross-Entity Signal Propagation and Alpha Discovery}}

\author{Lin Zhang\\[2pt]
\normalsize Cotality Data Science and AI Platform\thanks{Working paper. Comments
welcome: \texttt{linzhang1126@gmail.com}. All simulation code and
replication materials accompany this draft. The framework, results, and
any errors are the author's own. This draft: July 2026.}}

\date{July 2026}

\begin{document}

\maketitle

\begin{abstract}
\noindent
Financial information rarely affects a single company in isolation: earnings
surprises, capital-expenditure changes, supply constraints, and guidance
revisions propagate through networks of suppliers, customers, competitors,
and technology ecosystems. Traditional financial NLP estimates
document-level sentiment for the mentioned entity only, discarding the
cross-entity content of news. This paper develops a framework in which a
large language model (LLM) acts as a \emph{financial measurement engine}
that (i) converts unstructured documents into structured economic
\emph{state-change} events, (ii) extracts explicit and implicit economic
relationships to build a \emph{dynamic} financial knowledge graph, and
(iii) feeds a \emph{community-aware} signal-propagation mechanism in which
event signals diffuse more strongly within dynamically detected economic
communities than across them ($\lin>\lout$). We formalize the resulting
Community Information Surprise (CIS) and Propagated Information Surprise
(PIS) factors and derive the associated econometric tests. In a controlled
simulation of an economy with time-varying community structure---including
the mid-sample emergence of a new investment ecosystem---the feasible
estimator recovers the latent communities (mean NMI $\approx 0.86$), detects
the mid-sample ecosystem split as the network rewires, and the propagated
signal carries
incremental predictive power over the direct signal: across 20
replications, community-aware propagation attains the highest rank IC and
long--short Sharpe ratio among five nested benchmarks (sentiment, direct
LLM events, static-graph, dynamic-graph, and community-aware propagation),
with a median Fama--MacBeth $t$-statistic on the propagated signal of
$\approx 3.7$. A second, Russell-1000-calibrated configuration
($N=1{,}000$; sparser industry-group-scale graphs; large-cap volatility,
effect sizes, and heterogeneous news coverage) preserves the ordering at
realistic magnitudes---the propagated signal is priced in $80\%$ of
replications and the community gate is recovered with a difference
$t$-statistic of $3.6$---while quantifying the transaction-cost hurdle a
standalone implementation must clear. We also show analytically and empirically why
extraction noise in graph weights attenuates feasible propagation
coefficients, and we provide a complete point-in-time blueprint---data
sources, prompt design, and leakage controls---for a live Russell 1000
study.
\bigskip


\end{abstract}

\newpage
\tableofcontents
\newpage

\section{Introduction}
\label{sec:intro}

Modern asset markets digest an enormous flow of unstructured text:
earnings-call transcripts, regulatory filings, guidance revisions, trade
press, and supply-chain announcements. The dominant paradigm in financial
natural-language processing compresses each document $D_{i,t}$ associated
with firm $i$ at time $t$ into an entity-level score,
\begin{equation}
S_{i,t} \;=\; f_{\theta}\!\left(D_{i,t}\right),
\label{eq:trad-nlp}
\end{equation}
typically a sentiment polarity, and relates that score to the future
returns of the \emph{same} firm
\citep{tetlock2007giving,loughran2011liability,huang2023finbert}. Equation
\eqref{eq:trad-nlp} embeds a strong implicit assumption: the information in
a document is priced, if at all, only in the security of the entity it
mentions.

Real economies are not diagonal. A hyperscaler's decision to accelerate
AI-infrastructure capital expenditure is simultaneously a demand signal for
GPU vendors, semiconductor foundries, high-bandwidth-memory (HBM)
suppliers, advanced-packaging houses, optical-networking companies, and
power-infrastructure providers. A supply constraint at a lithography
monopolist is information about every fab that queues for its machines. A
large literature documents that markets absorb such cross-entity
information slowly: returns of economically linked firms predict each other
at horizons of weeks to months
\citep{cohen2008economic,menzly2010market,hou2007industry,hong2007industries},
consistent with theories of gradual information flow and limited attention
\citep{hong1999unified}.

This paper asks whether the two ingredients that have historically limited
the exploitation of cross-entity information---(i) the difficulty of
measuring \emph{events} rather than tone in text, and (ii) the difficulty
of maintaining an accurate, current map of \emph{who is economically
connected to whom}---can now be supplied by large language models, and
whether the resulting signals are incrementally priced. Our central
research question is:

\begin{quote}
\emph{Can LLM-extracted financial events generate predictive cross-entity
investment signals when propagated through dynamically detected financial
communities?}
\end{quote}

We make three conceptual moves. First, we reframe financial NLP as
\emph{latent economic state estimation}: the object of interest is not the
tone of a document but the \emph{innovation} it reveals about a firm's
economic state relative to the market's prior information set,
$\Delta \mathrm{State}_{i,t} = \mathrm{State}_{i,t} -
\E[\mathrm{State}_{i,t} \mid \F_{t-1}]$. An LLM that reads the current
document \emph{in the context of} prior documents and consensus
expectations acts as a measurement engine for this innovation, naturally
distinguishing genuinely new information from repeated positive boilerplate.

Second, we replace the static industry taxonomy with a \emph{dynamic
financial knowledge graph} $G_t=(V,E_t,W_t)$ whose edges (supplier,
customer, competitor, technology, capex-exposure, industry) and time-varying
weights are extracted by the LLM from the same document flow. On this graph
we run \emph{dynamic community detection}: latent economic communities
$\mathcal{C}_t$ are re-estimated as the graph evolves, allowing the
framework to watch a monolithic ``semiconductors'' cluster differentiate
into GPU-compute, HBM-memory, advanced-packaging, optical-networking, and
power-infrastructure ecosystems \emph{before} sector classifications catch
up.

Third, we propagate event signals across the graph with
\emph{community-aware} weights: a signal originating at firm $i$ reaches
firm $j$ with strength $w_{ij,t}\,\phi(C_{i,t},C_{j,t})$, where
$\phi=\lin$ inside a community and $\phi=\lout$ across communities, with
the maintained hypothesis $\lin>\lout$. Aggregating within communities
yields a Community Information Surprise (CIS) factor, and projecting CIS
back onto firms through their community exposures yields a firm-level
Propagated Information Surprise (PIS) factor---an investment signal for
firms about which \emph{no} new document was published.

Because a live LLM pipeline entangles model quality, data coverage, and
econometrics, we first validate the framework end-to-end in a controlled
simulation in which the truth is known and every real-world imperfection is
represented: the econometrician observes only noisy event signals (not the
true innovations), and only a noisy extracted graph (missing edges,
spurious edges, weight noise), while the true community structure changes
mid-sample. The simulation delivers four results. (1) Louvain-based
detection on the noisy extracted graph recovers the latent communities with
mean normalized mutual information of $0.87$ and detects the ecosystem
split as the network rewires over the following refresh cycles---the
detector's lag matching the economy's own gradual differentiation. (2) The diffusion mechanism is
directly visible in event studies: neighbours in the same community drift
for two to three weeks after a source event, while cross-community
neighbours barely move, and the feasible regression estimates
$\hat\lin>\hat\lout$ with the attenuation predicted by
errors-in-variables theory. (3) In Fama--MacBeth regressions the propagated
signal is priced \emph{incrementally} to the direct signal---the central
test $\beta_2\ne 0$---with median $t\approx3.7$ for community-aware
propagation across 20 replications, versus $3.4$ for uniform dynamic-graph
propagation and $1.1$ for a static graph frozen at the start of the sample.
(4) The economic ordering is monotone in rank IC---sentiment $<$ direct
events $<$ static propagation $<$ dynamic propagation $<$ community-aware
propagation---and community-aware propagation attains the highest
long--short Sharpe ratio, though portfolio-level differences among the
middle methods sit inside cross-replication noise; at 5 bps one-way costs
only the event- and propagation-based variants remain non-negative, with
the community-aware strategy best.

We then push the design to Russell 1000 scale. A second, fully
recalibrated configuration---$1{,}000$ firms, eighteen
industry-group-scale communities on a sparser extracted graph, large-cap
volatility, smaller event responses, and heterogeneous news coverage
drawn to mimic the concentration of documents in well-covered
names---re-runs the entire pipeline at the magnitudes a live study would
face. ICs compress by roughly $40\%$, exactly as a more efficient market
implies, yet the ordering survives and the central test remains
detectable: the propagated signal is priced at the $5\%$ level in $80\%$
of replications for community-aware propagation, the community gate is
recovered at $\hat\lin=5.4$ versus $\hat\lout=0.8$ bps (difference
$t=3.6$), and the exercise quantifies the transaction-cost
hurdle---gross Sharpe up to $1.16$ but no variant clearing 5 bps costs
as a standalone 10-day-rebalanced quintile strategy---that dictates how
the signal should be implemented (longer horizons, overlays,
low-coverage names).

Alongside this calibrated evaluation we provide the complete application
blueprint for the live study:
point-in-time data sources (earnings-call transcripts, SEC filings,
newswire with capture timestamps), the event and relationship extraction
prompts, leakage controls (LLM training-data cutoffs, as-of joins,
survivorship-free universes), and the evaluation protocol. We are explicit
that the simulation validates the \emph{machinery}---identification,
estimation, and backtesting under realistic measurement noise---while the
\emph{magnitude} of live alpha is an empirical question that depends on
extraction quality and market efficiency.

\paragraph{Contributions.} First, we reframe financial NLP as latent
economic state estimation rather than document sentiment classification,
and formalize the LLM as a measurement engine for state innovations.
Second, we introduce dynamic community detection into LLM-built financial
knowledge graphs, so that evolving economic ecosystems---not frozen sector
taxonomies---define the neighbourhoods over which information travels.
Third, we develop and test a community-aware propagation mechanism, with
the associated CIS/PIS factor construction and econometric identification
strategy, and characterize how graph-extraction noise attenuates feasible
propagation estimates. The broader hypothesis the framework operationalizes
is that alpha arises not only from discovering new information, but from
modelling \emph{how information travels} across latent economic communities
before prices fully incorporate it.

The paper proceeds as follows. Section~\ref{sec:background} reviews the
economic mechanisms and the methodological building blocks. Section~
\ref{sec:framework} develops the framework. Section~\ref{sec:econometrics}
states the hypotheses and estimation methodology. Section~\ref{sec:sim}
presents the simulation study. Section~\ref{sec:application} gives the
live-market application design. Section~\ref{sec:discussion} discusses
limitations, and Section~\ref{sec:conclusion} concludes. Appendices collect
propagation-operator properties, community-detection details, and prompt
templates.

\section{Mechanism Background and Related Literature}
\label{sec:background}

\subsection{Gradual information diffusion across economic links}
\label{sec:mech-diffusion}

The economic mechanism underlying the framework is \emph{gradual
information flow}. In \citet{hong1999unified}, information diffuses slowly
across a population of agents who each observe only part of the signal;
prices underreact at first and drift subsequently. The empirical
counterpart in a cross-firm setting is the lead--lag literature.
\citet{cohen2008economic} show that returns of a firm's principal
customers predict the firm's own future returns: a monthly long--short
strategy on ``customer momentum'' earns abnormal returns because investors
are inattentive to publicly disclosed supply-chain links.
\citet{menzly2010market} document analogous cross-predictability between
supplier and customer \emph{industries}; \citet{hou2007industry} shows
that within industries, big-firm returns lead small-firm returns,
consistent with information diffusing from the most-visible entities
outward; and \citet{hong2007industries} find that industry returns lead
the aggregate market by up to two months. \citet{acemoglu2012network}
provide the general-equilibrium foundation: in a production network,
idiosyncratic shocks to one sector propagate downstream and, when the
network is sufficiently asymmetric, do not wash out in the aggregate.
Recent evidence keeps the mechanism current:
\citet{hirshleifer2025news} show that the diffusion of news through
investor social networks generates return predictability of exactly the
gradual form the theory implies.

Three features of this literature motivate our design. First, the
propagation medium is an \emph{economic network}, not a statistical factor
structure: who buys from whom, who competes with whom, who depends on whose
technology. Second, the diffusion is \emph{slow}---weeks, not
minutes---because the linking information, while public, is costly to
collect and monitor; this is precisely the cost an LLM pipeline collapses.
Third, the strength of diffusion is \emph{heterogeneous} across link types
and economic proximity, which motivates community-dependent propagation
rather than uniform graph smoothing.

\subsection{From sentiment to events: text as measurement}
\label{sec:mech-text}

Text-based return prediction begins with dictionary methods:
\citet{tetlock2007giving} links pessimistic media tone to short-horizon
index returns, and \citet{loughran2011liability} show that
finance-specific word lists are essential because general-purpose
dictionaries misclassify financial language. Supervised transformers
improve the measurement further: FinBERT variants
\citep{araci2019finbert,huang2023finbert} classify financial sentences
with domain fine-tuning, and \citet{wu2023bloomberggpt} scale a
finance-specific LLM to 50B parameters. Most recently,
\citet{lopezlira2026chatgpt} show that a general-purpose LLM's assessment
of headline news predicts next-day returns, and \citet{chen2023expected}
find that LLM-based news representations subsume many classical text
predictors. The broader text-as-data programme supports treating text as
a measurement of economic state rather than tone: news topic attention
spans the macroeconomic state space \citep{bybee2024business}, scalable
textual factors carry pricing information \citep{cong2025textual}, and
LLM-extracted supply-chain risk measures are priced in the cross-section
\citep{fan2025measuring}. See \citet{kelly2023financial} for the
financial machine-learning landscape and \citet{nie2024survey} for a
survey of LLM applications in finance.

Two measurement problems survive even the best sentiment classifiers, and
both are addressed by event-level extraction with context. The first is
\emph{levels versus innovations}: a persistently optimistic company
generates positive tone every quarter, but only the \emph{unexpected}
component moves prices; a measurement of tone confounds the stock of old
(already-priced) information with the flow of new information. The second
is \emph{attribution}: a single document typically contains information
about multiple entities (a capex announcement is demand information for
suppliers), and entity-level scoring discards all off-diagonal content.
Our event-engineering layer (Section~\ref{sec:framework-events}) is
designed around exactly these two failures.

\subsection{Financial knowledge graphs}
\label{sec:mech-kg}

Structured firm-relationship data has traditionally come from disclosed
customer lists (Compustat segment files, as in \citealp{cohen2008economic}),
input--output tables (\citealp{menzly2010market,acemoglu2012network}), or
commercial supply-chain databases. These sources are accurate but sparse,
slow-moving, and biased toward disclosed, material relationships. Text
offers a complementary path. \citet{schwenkler2020network} extract a firm
network from news co-occurrence and show that news-implied links predict
comovement---the pre-LLM precursor of our extraction layer. With modern
language models the construction scales:
\citet{breitung2025global} build a global network of economically linked
firms from business descriptions and document cross-firm return
predictability along the learned links; \citet{li2024findkg} construct a
dynamic financial knowledge graph from news with an LLM and show that
graph-based trend detection identifies emerging themes (e.g., the AI
ecosystem) tradeably early; \citet{arun2025finreflectkg} develop agentic,
self-evaluating LLM pipelines for financial KG construction from filings;
and \citet{huang2026crossstock} use LLM-classified relationship types
from 10-Ks to build cross-stock reversal signals. Closest in spirit to
our propagation layer, \citet{chen2023chatgptgnn} feed ChatGPT-inferred
company graphs into a GNN for movement prediction on the Dow~30, and
\citet{wang2023knowledge} combine a knowledge graph, a GCN, and community
detection for stock prediction in an engineering setting.

Relative to this rapidly growing literature, our contribution is the
combination and its economics: firm-level \emph{events} measured as state
innovations (not descriptions or sentiment), a \emph{dynamic}
relationship graph whose \emph{community structure}---and its
evolution---is the central propagation medium, an a priori
identified linear operator with the community gate $\lin>\lout$ as a
testable hypothesis, and formal cross-sectional asset-pricing tests
(Fama--MacBeth incrementality, feasible $\lambda$ estimation with
attenuation theory) rather than movement-classification accuracy. To our
knowledge no prior paper joins all of these ingredients; the closest
works each hold one piece fixed---static description-based links
\citep{breitung2025global}, no community structure
\citep{chen2023chatgptgnn,huang2026crossstock}, or no LLM and no pricing
tests \citep{wang2023knowledge}.

\subsection{Community detection in networks}
\label{sec:mech-community}

A community is a set of nodes more densely connected internally than
externally. The canonical quality function is the modularity of
\citet{newman2004finding}: for a weighted graph with adjacency $A_{ij}$,
total edge weight $m=\tfrac12\sum_{ij}A_{ij}$ and (weighted) degrees
$k_i=\sum_j A_{ij}$,
\begin{equation}
Q \;=\; \frac{1}{2m}\sum_{ij}\left(A_{ij}-\frac{k_i k_j}{2m}\right)
\delta(C_i,C_j),
\label{eq:modularity}
\end{equation}
which compares the realized within-community weight to its expectation
under a degree-preserving null model. Exact modularity maximization is
NP-hard; the Louvain algorithm \citep{blondel2008fast} provides a fast,
greedy multi-level heuristic that scales to millions of edges and is the
workhorse we adopt. Alternatives include flow-based Infomap
\citep{rosvall2008maps}, statistically grounded stochastic block models
\citep{holland1983stochastic}, and---directly relevant here---multislice
modularity for time-dependent networks \citep{mucha2010community}, which
couples community assignments across adjacent time slices and has been
applied to financial correlation networks by \citet{bazzi2016community}.
See \citet{fortunato2010community} for a survey. In finance, correlation-based
networks already reveal economically meaningful hierarchical structure
\citep{mantegna1999hierarchical}; our graph is instead built from
\emph{stated economic relationships}, so its communities are ecosystems of
economic dependence rather than clusters of return comovement---an
important distinction, since comovement clusters cannot, by construction,
lead returns.

\subsection{Graph neural networks and learned propagation}
\label{sec:mech-gnn}

Our baseline propagation operator is a linear graph diffusion; its learned
generalization is a graph neural network. Graph convolutional networks
\citep{kipf2017semi} apply
$H^{(l+1)}=\sigma(\tilde D^{-1/2}\tilde A\tilde D^{-1/2}H^{(l)}W^{(l)})$;
GraphSAGE \citep{hamilton2017inductive} generalizes to inductive settings
with sampled neighbourhood aggregation; and graph attention networks
\citep{velickovic2018graph} let the model learn edge-specific propagation
weights,
\begin{equation}
h_j^{(l+1)}=\sigma\Big(\sum_{i\in\mathcal N(j)}\alpha_{ij}\,W h_i^{(l)}\Big),
\qquad
\alpha_{ij}=\mathrm{softmax}_i\big(a(Wh_i,Wh_j)\big),
\label{eq:gat}
\end{equation}
which is the natural vehicle for \emph{event-type-dependent} propagation: a
supply-constraint event should travel along supplier edges with a different
sign and speed than a demand-acceleration event travels along customer
edges. A substantial applied literature already uses relational GNNs for
stock prediction---temporal relational ranking over industry/knowledge
graphs \citep{feng2019temporal}, hierarchical attention over multi-type
corporate relations \citep{kim2019hats}, and temporal-heterogeneous GNNs
on dynamically regenerated company graphs \citep{xiang2022temporal}. We
use the linear, community-aware operator as the identifiable,
interpretable core, and view the GAT extension
(Section~\ref{sec:framework-prop}) as the learned refinement to be trained
once live data volume permits.

\section{The Framework}
\label{sec:framework}

The framework is a pipeline
\begin{equation*}
\text{LLM}
\rightarrow
\text{Event Engineering}
\rightarrow
\text{Dynamic Knowledge Graph}
\rightarrow
\text{Community Detection}
\rightarrow
\text{Signal Propagation}
\rightarrow
\text{Alpha Tests},
\end{equation*}
formalized on a dynamic graph
$G_t=(V,E_t,W_t)$ with $V$ the investable universe ($|V|=N$), $E_t$ the
set of economic relationships active at $t$, and
$W_t=[w_{ij,t}]\in\R_{+}^{N\times N}$ their time-varying strengths.

\subsection{LLM-based financial event engineering}
\label{sec:framework-events}

For each firm--document pair the extraction agent produces a structured
event tuple
\begin{equation}
e_{i,t}=(z,\,m,\,h,\,n,\,c),
\label{eq:event-tuple}
\end{equation}
with $z$ an event category (demand, capex, pricing, inventory, supply
constraint, guidance, regulatory, management), $m\in\R$ a signed direction
and magnitude, $h$ an expected economic horizon, $n\in[0,1]$ an information
novelty score, and $c\in[0,1]$ extraction confidence. The scalar signal
used downstream is the confidence- and novelty-weighted magnitude, which we
interpret as a noisy measurement of the firm's latent state innovation:
\begin{equation}
s_{i,t} \;=\; f_\theta\!\left(D_{i,t},\,D_{i,t-1},\,\F_{t-1}\right)
\;=\; \Delta \mathrm{State}_{i,t} + \eta_{i,t},
\qquad
\Delta \mathrm{State}_{i,t}
= \mathrm{State}_{i,t}-\E\!\left[\mathrm{State}_{i,t}\,\middle|\,\F_{t-1}\right],
\label{eq:deltastate}
\end{equation}
where $\eta_{i,t}$ is measurement noise with variance decreasing in $c$.
The conditioning on $D_{i,t-1}$ and $\F_{t-1}$ (prior documents, consensus
expectations, guidance history) is what makes $s_{i,t}$ an
\emph{innovation} measure: a firm repeating last quarter's positive outlook
generates $s_{i,t}\approx 0$, while a small negative surprise from a
chronically optimistic firm generates $s_{i,t}<0$ even though the document's
tone is positive. This is the operational content of ``latent economic
state estimation'' and the key difference from sentiment scoring
\eqref{eq:trad-nlp}.

\subsection{Dynamic knowledge-graph construction}
\label{sec:framework-kg}

The same document flow is mined for typed relationships
\begin{equation*}
\rho \in \{\textit{Supplier},\ \textit{Customer},\ \textit{Competitor},\
\textit{Technology},\ \textit{CapExExposure},\ \textit{Industry}\},
\end{equation*}
each with a direction, an intensity, and a confidence. Edge weights evolve
as new evidence arrives and old evidence decays:
\begin{equation}
w_{ij,t} \;=\;
g\!\left(\mathrm{Relationship}_{ij,t},\ \mathrm{LLMContext}_{ij,t},\
\mathrm{MarketData}_{ij,t}\right),
\label{eq:edge-weight}
\end{equation}
in practice an exponentially weighted evidence count per relationship type,
optionally corroborated by market data (e.g., co-movement at announcement
times) but never dominated by it, so that the graph retains its ability to
\emph{lead} returns. The graph is refreshed at a fixed cadence (monthly in
our implementation). Because extraction is imperfect, the observed graph
$\widehat W_t$ differs from the true economic network $W_t$ by missing
edges, spurious edges, and weight noise; Section~\ref{sec:sim} models this
explicitly, and Proposition~\ref{prop:attenuation} quantifies its effect on
estimated propagation.

\subsection{Dynamic community detection}
\label{sec:framework-community}

At each refresh date the framework partitions the graph into communities
$\mathcal C_t=\{C_{1,t},\dots,C_{K_t,t}\}$ by maximizing modularity
\eqref{eq:modularity} on $\widehat W_t$ with the Louvain algorithm
\citep{blondel2008fast}; the number of communities $K_t$ is not fixed
a priori. Allowing $\mathcal C_t$ to vary over time is the central
methodological extension: a traditional ``semiconductor'' community may
differentiate into economically distinct AI-era ecosystems,
\begin{equation*}
\textit{GPU compute}\ \rightarrow\ \textit{HBM memory}\ \rightarrow\
\textit{advanced packaging}\ \rightarrow\ \textit{optical networking}\
\rightarrow\ \textit{power infrastructure},
\end{equation*}
and the framework should detect such splits from the changing edge
structure \emph{before} they are reflected in static sector
classifications. For temporal stability one may add inter-slice coupling
\citep{mucha2010community}; we found simple per-slice Louvain with a fixed
random seed sufficient in simulation, with label matching across slices for
diagnostics only (the propagation operator depends on the partition, not on
label names).

\subsection{Graph-based signal propagation}
\label{sec:framework-prop}

Let $s_t\in\R^N$ stack the (decayed) event signals. The baseline
propagation operator is a $K$-step damped diffusion,
\begin{equation}
\tilde s_t \;=\; \sum_{k=1}^{K}\gamma^k P_t^{\,k}\, s_t,
\qquad 0<\gamma<1,
\label{eq:baseline-prop}
\end{equation}
where $P_t$ is a normalization of $\widehat W_t$ (we scale by mean degree;
row-stochastic and symmetric normalizations are alternatives, and all
downstream statistics are invariant to global scale because signals are
cross-sectionally standardized). The direct ($k=0$) term is deliberately
excluded from $\tilde s_t$ and carried as a separate regressor so that the
incremental content of \emph{neighbours'} information can be tested
(Section~\ref{sec:econometrics}). Appendix~\ref{app:operator} gives
convergence conditions ($\gamma\,\rho(P_t)<1$) and the interpretation of
\eqref{eq:baseline-prop} as a truncated Neumann series of the resolvent
$(I-\gamma P_t)^{-1}$.

The community-aware operator reweights edges by community co-membership
before normalizing:
\begin{equation}
\tilde s_{j,t}\;=\;\sum_{i} w_{ij,t}\,
\phi\!\left(C_{i,t},C_{j,t}\right) s_{i,t},
\qquad
\phi(C_i,C_j)=
\begin{cases}
\lin, & C_i=C_j,\\[2pt]
\lout, & C_i\neq C_j,
\end{cases}
\label{eq:comm-prop}
\end{equation}
with maintained hypothesis $\lin>\lout$. The pair $(\lin,\lout)$ can be
fixed a priori (we use $(1,0.3)$), estimated by the pooled regression of
Section~\ref{sec:econometrics}, or generalized to a full matrix
$\phi(k,k')$ over community pairs. Beyond down-weighting genuinely weak
cross-community diffusion, \eqref{eq:comm-prop} has a second, purely
statistical benefit: extraction errors (spurious edges) are
disproportionately cross-community, so community-aware damping filters
graph noise.

The learned generalization replaces \eqref{eq:comm-prop} with a graph
attention network \eqref{eq:gat} in which the attention logits take as
inputs the event tuple \eqref{eq:event-tuple} (so that supply-constraint
events propagate differently from demand events), the edge type $\rho$,
and the community co-membership indicator. We view identification of the
linear operator as a prerequisite: the GAT nests it and should be trained
only when live data volume supports it.

\subsection{Community and propagated information surprise factors}
\label{sec:framework-cis}

Aggregating state innovations within a community gives the
\emph{Community Information Surprise}:
\begin{equation}
\mathrm{CIS}_{k,t}\;=\;\sum_{i\in C_{k,t}}\omega_{i,t}\,
\Delta \mathrm{State}_{i,t},
\qquad \omega_{i,t}\ \text{attention weights},\ \textstyle\sum_i
\omega_{i,t}=1,
\label{eq:cis}
\end{equation}
with $\omega$ proportional to size, centrality, or extraction confidence
(equal weights in our implementation). Projecting community surprises back
onto firms through exposures yields the \emph{Propagated Information
Surprise}:
\begin{equation}
\mathrm{PIS}_{j,t}\;=\;\sum_{k}\mathrm{Exposure}_{j,k,t}\times
\mathrm{CIS}_{k,t},
\qquad
\mathrm{Exposure}_{j,k,t}
=\frac{\sum_{i\in C_{k,t}} w_{ij,t}\,\phi(C_{i,t},C_{j,t})}
{\sum_{k'}\sum_{i\in C_{k',t}} w_{ij,t}\,\phi(C_{i,t},C_{j,t})}.
\label{eq:pis}
\end{equation}
The economically important property of \eqref{eq:pis} is that
$\mathrm{PIS}_{j,t}$ can be large when \emph{no} document mentions firm
$j$: the information originates elsewhere in $j$'s economic network. PIS
is a smoothed, community-level counterpart of the edge-level propagation
\eqref{eq:comm-prop}; the two are correlated but not identical, and both
enter the horse-race in Section~\ref{sec:sim}.

\section{Hypotheses and Econometric Methodology}
\label{sec:econometrics}

\subsection{Hypotheses}

Let $r_{j,t+h}$ denote firm $j$'s return over $(t,t+h]$ and $\F_t$ the
market information set. The framework's claims are:

\begin{hypothesis}[Cross-entity information content]
\label{hyp:cross}
For economically connected $i\ne j$,
$\E[r_{j,t+h}\mid \F_t, s_{i,t}] \neq \E[r_{j,t+h}\mid \F_t]$
at horizons $h$ of days to weeks.
\end{hypothesis}

\begin{hypothesis}[Community-gated propagation]
\label{hyp:community}
Signal transmission is stronger within than across communities:
$\Prob(\alpha_{j,t+h}>0 \mid s_{i,t}, C_{i,t}=C_{j,t})
>\Prob(\alpha_{j,t+h}>0 \mid s_{i,t}, C_{i,t}\neq C_{j,t})$;
equivalently $\lin>\lout$ in \eqref{eq:comm-prop}.
\end{hypothesis}

\begin{hypothesis}[Incremental pricing]
\label{hyp:incremental}
The propagated signal predicts returns incrementally to the direct signal:
$\beta_2\ne0$ in \eqref{eq:fm} below.
\end{hypothesis}

\subsection{Cross-sectional pricing tests}

The central regression is estimated cross-sectionally each day and
aggregated in the \citet{fama1973risk} manner:
\begin{equation}
r_{i,t+1:t+h}
=\alpha_t
+\beta_{1,t}\,\mathrm{Direct}_{i,t}
+\beta_{2,t}\,\mathrm{Propagated}_{i,t}
+\Gamma_t' X_{i,t}
+\epsilon_{i,t+h},
\label{eq:fm}
\end{equation}
with $X_{i,t}$ feasible controls (market beta estimated on a trailing
window, short-term reversal, momentum) and all signals cross-sectionally
standardized. The test of
Hypothesis~\ref{hyp:incremental} is $H_0:\E[\beta_{2,t}]=0$ against
$H_A:\E[\beta_{2,t}]\ne0$. Because $h$-day returns overlap, time-series
means of $\{\hat\beta_{2,t}\}$ are assessed with \citet{newey1987simple}
standard errors with $h$ lags.

The direct estimator of Hypothesis~\ref{hyp:community} is the pooled
edge-level regression
\begin{equation}
r_{j,t+1:t+h}
=a+\lambda_{g}\,\big(\widehat w_{ij,t}\, s_{i,t}\big)
+u_{ij,t},
\qquad
g=\begin{cases}
\text{in}, & \widehat C_{i,t}=\widehat C_{j,t},\\
\text{out}, & \widehat C_{i,t}\ne\widehat C_{j,t},
\end{cases}
\label{eq:lambda-reg}
\end{equation}
run over source events $(i,t)$ and their graph neighbours $j$ (excluding
neighbours with own contemporaneous events), with the difference
$\hat\lin-\hat\lout$ as the statistic. Both the regressor (observed
weights, noisy signals) and the conditioning (detected communities) are
feasible, so \eqref{eq:lambda-reg} is an estimator an econometrician could
run on live data. It is, however, attenuated by measurement error:

\begin{proposition}[Attenuation from graph-extraction noise]
\label{prop:attenuation}
Suppose the true per-edge response is
$\E[r_{j,t+1:t+h}\mid u_{i,t}, w_{ij,t}]=\lambda\, w_{ij,t} u_{i,t}$ and
the econometrician observes $\widehat w_{ij,t}=w_{ij,t}+e_{ij}$,
$s_{i,t}=u_{i,t}+\eta_{i,t}$ with independent, mean-zero errors. Then the
OLS slope of $r_{j}$ on $\widehat w_{ij}s_{i}$ converges to
\begin{equation}
\lambda\cdot
\underbrace{\frac{\sigma_u^2}{\sigma_u^2+\sigma_\eta^2}}_{\text{signal noise}}
\cdot
\underbrace{\frac{\E[w^2]}{\E[w^2]+\sigma_e^2}}_{\text{weight noise}}
\;<\;\lambda .
\label{eq:attenuation}
\end{equation}
\end{proposition}

\noindent
The proof is the standard errors-in-variables calculation applied to the
product regressor (Appendix~\ref{app:operator}). Equation
\eqref{eq:attenuation} matters in practice: it implies that feasible
estimates of $\lin,\lout$ understate true diffusion, that improving
extraction quality has a first-order effect on measured propagation, and it
is verified quantitatively in our simulation
(Section~\ref{sec:sim-mechanism}).

\subsection{Portfolio evaluation}

Signal quality is summarized by the daily cross-sectional Spearman rank
information coefficient
$\mathrm{IC}_t^{(h)}=\mathrm{corr}_{\mathrm{rank}}
(\mathrm{signal}_{\cdot,t},\,r_{\cdot,t+1:t+h})$, its decay profile in
$h$, and the information ratio
$\mathrm{ICIR}=\overline{\mathrm{IC}}/\sigma(\mathrm{IC})$. Implementable
performance is assessed with quintile long--short portfolios: equal-weight
top-minus-bottom quintile, rebalanced every 10 days, with proportional
one-way transaction costs of 5 and 10 basis points applied to two-sided
turnover; we report annualized gross and net returns, Sharpe ratios,
maximum drawdown, and turnover.

\subsection{Point-in-time discipline}

Every quantity indexed by $t$ is computed from information available at
$t$: signals use documents up to $t$; the dynamic graph and communities use
the extraction as of the most recent refresh $\le t$; the ``static graph''
benchmark is frozen using only the first months of the sample (no
full-sample averaging, which would leak future relationships); forward
returns start at $t+1$. In live applications this discipline extends to
LLM training-cutoff controls and as-of database joins
(Section~\ref{sec:application-pit}).

\section{Simulation Study}
\label{sec:sim}

A live pipeline confounds three questions: is the machinery sound, is the
LLM measurement good, and is the market inefficient enough to leave the
alpha unarbitraged? The simulation isolates the first question while
representing the failures of the second: the econometrician sees only noisy
signals on event days and a noisy extracted graph, never the truth.

\subsection{Data generating process}
\label{sec:sim-dgp}

We simulate $N=300$ firms over $T=750$ trading days (three years).
Table~\ref{tab:params} collects all parameters.

\paragraph{Latent communities.} Firms begin in $K_0=6$ communities of
heterogeneous size. At $t=375$ the largest community splits in two---the
``ecosystem emergence'' event---and $5\%$ of firms migrate between
communities at random dates, so the partition $\mathcal C_t$ is genuinely
dynamic.

\paragraph{Dynamic graph.} Conditional on current memberships, edges follow
a stochastic block model \citep{holland1983stochastic} with within/cross
edge probabilities $p_{\mathrm{in}}=0.18$, $p_{\mathrm{out}}=0.012$ and
weights drawn from community-dependent uniforms. The graph refreshes every
21 days with edge survival probability $0.85$, edge birth to maintain
block densities, and AR(1) weight persistence $0.9$---so the network has
real turnover and the split reshapes connectivity over a few months rather
than instantaneously. The edge process is burned in for ten pre-sample
rebuilds so that the day-0 graph already sits at its stationary density;
without this, early graphs would be mechanically denser and the
static-graph benchmark (extracted early in the sample) would inherit a
density advantage unrelated to the staleness effect we want to isolate.

\paragraph{Events and returns.} Each firm-day has an event with probability
$0.05$; events carry innovations $u_{i,t}\sim\mathcal N(0,1)$. Returns
follow a one-factor structure plus event responses: the own firm responds
by $c_{\mathrm{own}}=50$ bps per unit shock over days 1--3, and each
neighbour responds \emph{per edge} by $c_{\mathrm{in}}=12$ bps (same
community) or $c_{\mathrm{out}}=3$ bps (different community) per unit
shock and unit edge weight, spread over days 1--12 by a geometric kernel.
The 4:1 ratio $c_{\mathrm{in}}/c_{\mathrm{out}}$ is the structural
counterpart of $\lin>\lout$; idiosyncratic volatility of $1.4\%$/day makes
individual responses small relative to noise, as in real data.

\paragraph{Measurement layer.} The econometrician observes (i) event
signals $s_{i,t}=u_{i,t}+\eta_{i,t}$, $\eta\sim\mathcal N(0,0.5^2)$, only
on event days; (ii) a level-based ``sentiment'' that measures the
persistent state level (persistence $0.95$) with noise---so it confounds
old and new information by construction; and (iii) an extracted graph
$\widehat W_t$ that misses $20\%$ of true edges, adds spurious edges equal
to $10\%$ of the true count, and perturbs weights with $\mathcal
N(0,0.15^2)$ noise. Nothing downstream ever touches $u$, $W$, or the true
membership except for oracle diagnostics that are clearly labelled as such.

\paragraph{Methods compared.} Five nested signals, all strictly
point-in-time: \textbf{M1} level sentiment (direct only); \textbf{M2}
direct LLM event signal (exponentially decayed stock of $s_{i,t}$,
half-life 5 days); \textbf{M3} = M2 plus propagation
\eqref{eq:baseline-prop} on a static graph frozen after the first six
refreshes; \textbf{M4} = M2 plus propagation on the current extracted
graph; \textbf{M5} = M2 plus community-aware propagation
\eqref{eq:comm-prop} with Louvain communities re-detected at each refresh
and $(\lin,\lout)=(1,0.3)$. Propagation uses $K=2$, $\gamma=0.35$
throughout.


\subsection{Community recovery and ecosystem-split detection}
\label{sec:sim-community}

\begin{figure}[t!]
\centering
\includegraphics[width=\textwidth]{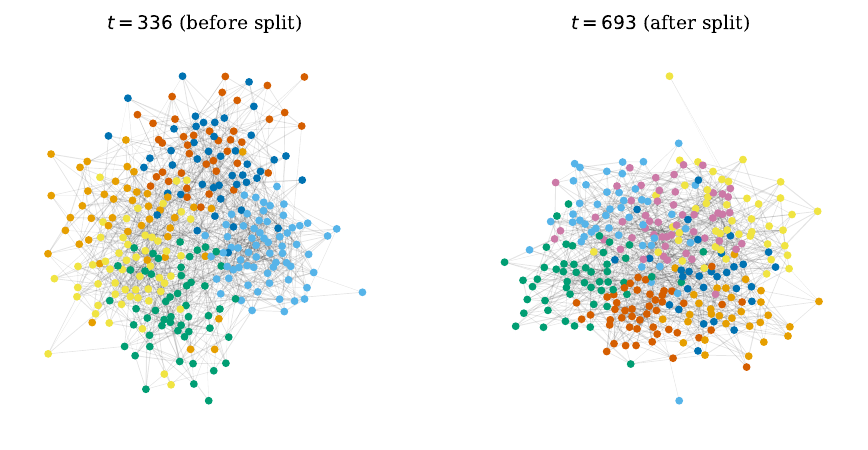}
\caption{\textbf{Extracted knowledge graph and detected communities.}
Spring-layout snapshots of the observed (noisily extracted) graph
$\widehat W_t$ in a representative replication, before (left) and after
(right) the true ecosystem split at $t=375$. Node colours are the
communities detected by weighted Louvain on $\widehat W_t$ at that
refresh date; edges drawn with opacity proportional to weight.}
\label{fig:network}
\end{figure}

Figure~\ref{fig:network} shows the extracted graph of a representative
replication before and after the ecosystem split, with node colours given
by the \emph{detected} (Louvain) communities. The block structure is
clearly recovered from the noisy extraction, and after $t=375$ the former
largest community appears as two distinct clusters.

\begin{figure}[t!]
\centering
\includegraphics[width=0.72\textwidth]{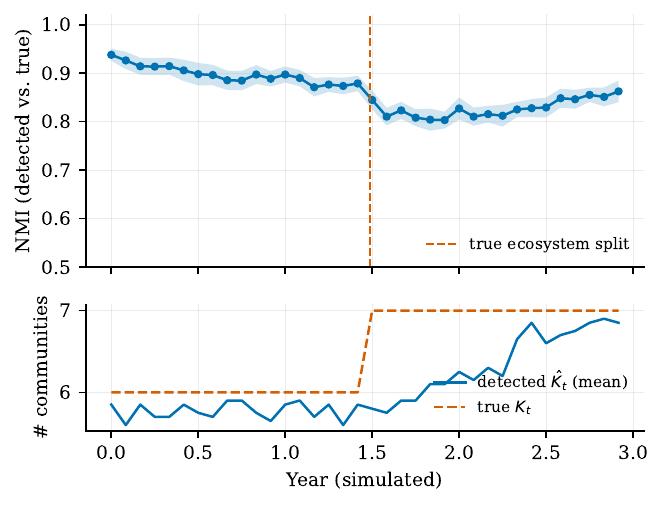}
\caption{\textbf{Dynamic community recovery.} Top: normalized mutual
information between Louvain communities detected on the extracted graph
and the true membership, at each monthly refresh; mean across 20
replications with $95\%$ band. Bottom: mean detected number of communities
$\hat K_t$ against the true $K_t$. The dashed vertical line marks the true
ecosystem split at $t=375$.}
\label{fig:nmi}
\end{figure}

Figure~\ref{fig:nmi} quantifies recovery over time. Across replications,
the mean normalized mutual information between detected and true
partitions is $0.86$, despite the extractor missing $20\%$ of true edges
and adding spurious ones. The lower panel shows the detected number of
communities against the true $K_t$. Detection of the split is
\emph{gradual}, and instructively so: memberships change at $t=375$, but
the edges connecting the two halves---inherited from their common
past---decay only at the graph's own turnover rate (edge survival $0.85$
per monthly rebuild, a half-life of roughly four months). Louvain tracks
the \emph{graph}, so the mean detected count rises from six toward seven
over the subsequent months as the legacy edges die and the new boundary
sharpens, while the NMI dips at the split and then recovers. This is the
correct behaviour for live markets, where an emerging ecosystem also
differentiates over quarters rather than overnight: the detector's lag is
the economy's lag, not an artifact of the method. The same panel shows a
mild conservatism before the split (mean $\hat K_t\approx5.7$ against a
true $6$), reflecting Louvain's occasional merging of the two
smallest communities---a known resolution-limit effect
\citep{fortunato2010community}.

\subsection{The diffusion mechanism: event studies and $\lambda$ estimates}
\label{sec:sim-mechanism}

\begin{figure}[t!]
\centering
\includegraphics[width=0.6\textwidth]{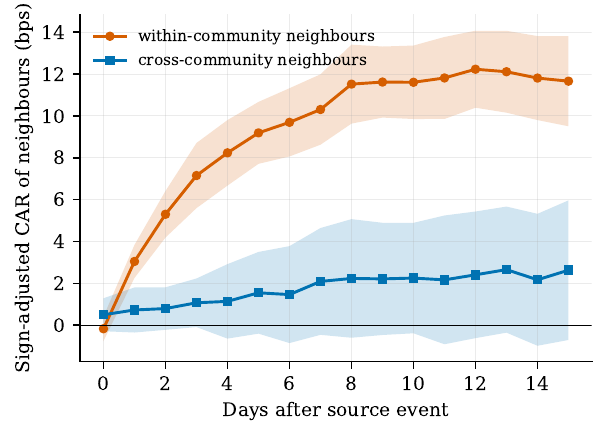}
\caption{\textbf{Cross-entity diffusion after source events.}
Sign-adjusted, market-adjusted cumulative average returns of graph
neighbours following source events with $|u_{i,t}|\ge0.75$, split by
true community co-membership; neighbours with own events inside the
window are excluded. Mean across 20 replications; shaded bands are
$95\%$ confidence intervals from cross-replication dispersion.}
\label{fig:eventstudy}
\end{figure}

Figure~\ref{fig:eventstudy} verifies the propagation mechanism the way an
empiricist would: an event study of neighbours' returns around source
events. For every event with $|u_{i,t}|\ge0.75$ we track the sign-adjusted,
market-adjusted cumulative return of the source firm's graph neighbours
(excluding neighbours with their own contemporaneous events), split by
whether the neighbour shares the source's community. Within-community
neighbours drift steadily to $9$ basis points by day 5 and about $12$
basis points by day 15---close to the true per-edge mass
$c_{\mathrm{in}}\times\bar w_{\mathrm{in}}\times\E|u|$---while
cross-community neighbours show only the small drift implied by
$c_{\mathrm{out}}$. The gradualness of
the within-community curve is the delayed information diffusion that
propagation-based signals monetize.

\begin{figure}[t!]
\centering
\includegraphics[width=0.38\textwidth]{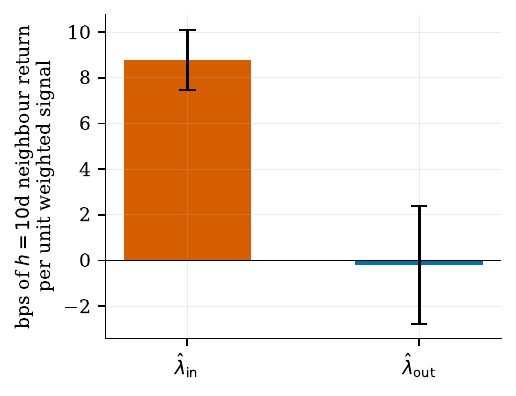}
\caption{\textbf{Feasible estimates of the community gate.} Pooled
edge-level regression \eqref{eq:lambda-reg} of neighbours'
10-day-forward market-adjusted returns on the weighted source signal,
split by \emph{detected} community co-membership. Bars are means across
20 replications; whiskers are $95\%$ confidence intervals. True per-edge
responses are $c_{\mathrm{in}}=12$ and $c_{\mathrm{out}}=3$ bps;
the shortfall is the measurement-error attenuation of
Proposition~\ref{prop:attenuation}.}
\label{fig:lambda}
\end{figure}

The regression counterpart is the feasible estimator
\eqref{eq:lambda-reg}, which uses only observables: extracted weights
$\widehat w_{ij,t}$, noisy event-day signals $s_{i,t}$, and
\emph{detected} communities. Pooling across replications
(Figure~\ref{fig:lambda}), $\hat\lin=8.8$ bps (s.e.\ $0.7$) and
$\hat\lout=-0.2$ bps (s.e.\ $1.3$) per unit of weighted signal over the
following ten days, with a difference $t$-statistic of $5.7$: the
community gate $\lin>\lout$ of Hypothesis~\ref{hyp:community} is
recovered from feasible data. Note the attenuation: the true
within-community response is $c_{\mathrm{in}}=12$ bps, so the feasible
estimate is $\approx73\%$ of the truth---quantitatively in line with
Proposition~\ref{prop:attenuation}, whose signal-noise and weight-noise
factors predict $\approx77\%$ before accounting for missing and
spurious edges and kernel truncation. Extraction quality is thus a
first-order determinant of \emph{measured} propagation, an effect that
carries over directly to live LLM pipelines.

\subsection{Predictive content and horizon structure}
\label{sec:sim-ic}

\begin{figure}[t!]
\centering
\includegraphics[width=0.62\textwidth]{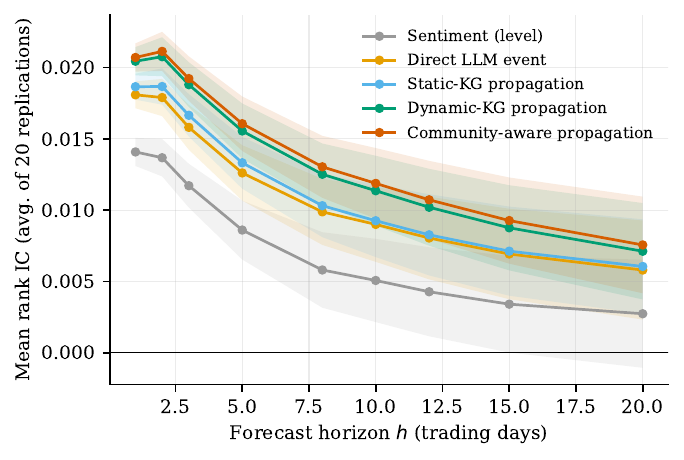}
\caption{\textbf{IC decay.} Mean rank IC of each method against forward
returns over horizons $h=1$--$20$ trading days; averages across 20
replications with $95\%$ bands.}
\label{fig:icdecay}
\end{figure}

Table~\ref{tab:ic} reports the primary-horizon ($h=5$) rank IC of the five
methods, and Figure~\ref{fig:icdecay} the full IC decay profiles. Three
features stand out. First, the ordering is monotone in the information
actually used: sentiment $<$ direct events $<$ static propagation $<$
dynamic propagation $<$ community-aware propagation, with mean
$h{=}5$ ICs of $0.009$, $0.013$, $0.013$, $0.016$, and $0.016$, and
ICIRs rising from $0.15$ to $0.28$. Second, the propagation-based signals
decay more slowly: the direct signal's IC falls from $0.018$ at $h=1$ to
$0.006$ at $h=20$ as the own-firm response is quickly priced, while
dynamic and community-aware propagation retain an IC near $0.008$ at
$h=20$---about a third higher than the direct signal---reflecting the
longer diffusion kernel of neighbour responses; this is the horizon at
which cross-entity strategies have capacity. Third, the level-based
sentiment signal is dominated at every horizon by the innovation-based
direct signal even though both are built from the same underlying events:
measuring $\Delta$State rather than tone raises the $h{=}5$ IC by nearly
half before any graph information is used, with the gap widening at
longer horizons as the stale components of the level provide no incremental
forecast.

\begin{table}[t!]
\centering
\caption{\textbf{Rank IC at the primary horizon ($h=5$).} Cross-sectional
Spearman correlation between each signal and 5-day-forward returns,
averaged over evaluation days; entries are means across 20 replications
(cross-replication standard deviation in parentheses for the mean IC and
ICIR). The $t$-statistic uses an overlapping-window correction.}
\label{tab:ic}
\medskip
\small
\begin{tabular}{lcccc}
\toprule
Method & Mean IC & IC vol & ICIR & $t$-stat \\
\midrule
Sentiment (level) & 0.0086 (0.0047) & 0.0573 & 0.149 (0.080) & 1.73 \\
Direct LLM event & 0.0126 (0.0044) & 0.0572 & 0.219 (0.072) & 2.53 \\
Static-KG propagation & 0.0133 (0.0041) & 0.0567 & 0.234 (0.070) & 2.70 \\
Dynamic-KG propagation & 0.0156 (0.0044) & 0.0568 & 0.273 (0.073) & 3.15 \\
Community-aware propagation & 0.0161 (0.0044) & 0.0568 & 0.282 (0.073) & 3.26 \\
\bottomrule
\end{tabular}
\end{table}

\subsection{Incremental pricing of the propagated signal}
\label{sec:sim-fm}

Table~\ref{tab:fm} contains the central test. Specification (1) confirms
that even the level-based sentiment carries information (the level
overlaps recent innovations). Specification (2) shows the direct
$\Delta$State signal is strongly priced (median $t=5.9$).
Specifications (3)--(5) add the propagated signal built on the static,
dynamic, and community-aware graphs: the propagated
coefficient---$\beta_2$ in \eqref{eq:fm}---is insignificant on the graph
frozen at the start of the sample (median $t=1.1$; $H_0$ rejected in
only $20\%$ of replications), but strongly significant on the dynamic
graph (median $t=3.4$; rejection rate $85\%$) and the community-aware
graph (median $t=3.7$; rejection rate $90\%$), while the
direct coefficient is essentially unchanged. This is
Hypothesis~\ref{hyp:incremental}: neighbours' events predict a firm's
return \emph{beyond} the firm's own news. The static-graph failure is
informative rather than mechanical---the frozen graph is extracted from
the same process and is accurate early in the sample; its propagated
signal dies because relationships and communities \emph{change}, which is
precisely the case for dynamic knowledge graphs. Specification (6) adds
the community-level PIS factor \eqref{eq:pis} on top of edge-level
propagation; it enters with a small positive coefficient and is
insignificant, indicating that once edge-level community-aware
propagation is included, the coarser community-average factor is
subsumed. PIS remains valuable as a standalone signal where edge-level
propagation is unavailable---most notably for firms with sparse direct
coverage---but it is not an independent source of alpha on top of
\eqref{eq:comm-prop} in this calibration.

\begin{table}[t!]
\centering
\caption{\textbf{Fama--MacBeth regressions of 5-day-forward returns (in
percent) on standardized signals.} Daily cross-sectional OLS with
controls for market beta (estimated on a trailing 60-day window against
the equal-weight market; not shown), 5-day reversal, and 60-day momentum;
time-series means of daily coefficients with \citet{newey1987simple}
$t$-statistics (5 lags). Coefficients are cross-replication means (20
replications); parenthesized $t$-statistics and significance stars are
keyed to the cross-replication \emph{median} of the per-replication
$t$-statistics, and the rejection-rate row---the fraction of replications
in which $H_0:\beta_2=0$ is rejected at the $5\%$ level---is the primary
inference. $^{*}$, $^{**}$, $^{***}$: median $|t|$ exceeds $1.64$,
$1.96$, $2.58$.}
\label{tab:fm}
\medskip
\small
\begin{tabular}{lcccccc}
\toprule
 & (1) & (2) & (3) & (4) & (5) & (6) \\
\midrule
Sentiment$_{i,t}$ & 0.049$^{***}$ &  &  &  &  &  \\
 & (3.64) &  &  &  &  &  \\[2pt]
Direct$_{i,t}$ &  & 0.072$^{***}$ & 0.071$^{***}$ & 0.068$^{***}$ & 0.067$^{***}$ & 0.067$^{***}$ \\
 &  & (5.86) & (5.68) & (5.55) & (5.47) & (5.42) \\[2pt]
Propagated$_{i,t}$ &  &  & 0.013 & 0.040$^{***}$ & 0.043$^{***}$ & 0.042$^{***}$ \\
 &  &  & (1.14) & (3.36) & (3.72) & (3.08) \\[2pt]
PIS$_{i,t}$ &  &  &  &  &  & 0.004 \\
 &  &  &  &  &  & (0.35) \\[2pt]
Reversal$_{i,t}$ & 0.004 & 0.002 & 0.002 & 0.002 & 0.002 & 0.001 \\
 & (0.52) & (0.41) & (0.36) & (0.37) & (0.36) & (0.40) \\[2pt]
Momentum$_{i,t}$ & -0.006 & -0.005 & -0.006 & -0.005 & -0.005 & -0.006 \\
 & (-0.60) & (-0.58) & (-0.56) & (-0.63) & (-0.61) & (-0.66) \\[2pt]
\midrule
Rej.\ rate $H_0\!:\beta_2{=}0$ & -- & -- & 20\% & 85\% & 90\% & 95\% \\
Avg.\ $R^2$ & 3.2\% & 3.2\% & 3.5\% & 3.5\% & 3.5\% & 3.9\% \\
\bottomrule
\end{tabular}
\end{table}

\subsection{Portfolio performance}
\label{sec:sim-portfolios}

\begin{figure}[t!]
\centering
\includegraphics[width=0.62\textwidth]{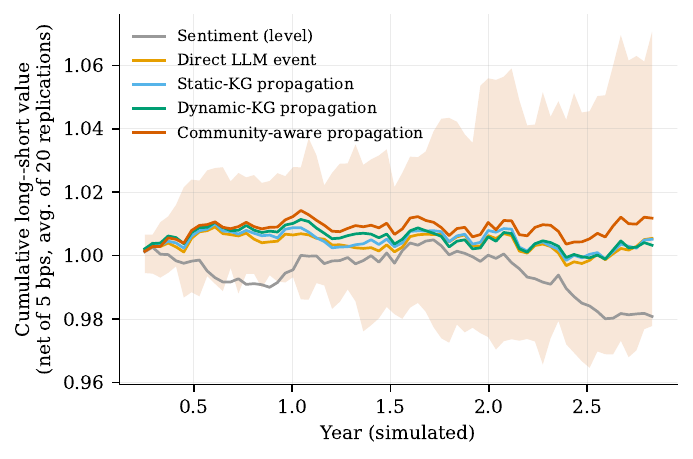}
\caption{\textbf{Cumulative long--short performance net of 5 bps costs.}
Average cumulative portfolio value across 20 replications; the shaded
band is the interquartile range for the community-aware method.}
\label{fig:cumret}
\end{figure}

Table~\ref{tab:portfolio} translates the ICs into quintile long--short
portfolios (10-day rebalancing), and Figure~\ref{fig:cumret} shows average
cumulative net-of-cost paths. The portfolio evidence is cleanly ordered at
the extremes and honest about the middle: sentiment earns a gross Sharpe
of $0.45$, the three event-based methods cluster at $0.75$, and
community-aware propagation is highest at $0.83$---an $11\%$ improvement
over the direct signal at the sort level. That the IC advantage of
propagation compresses in quintile sorts is expected: sorts are coarse,
the composite signal is dominated by its direct component in the tails,
and the cross-replication dispersion of Sharpe ratios ($\approx0.5$)
exceeds the method gaps for the middle three methods---only sentiment
versus the rest and community-aware at the top separate reliably. The
Fama--MacBeth evidence in Table~\ref{tab:fm}, which uses the full
cross-section rather than tail buckets, is the sharper lens on
incrementality. Costs matter for all variants: turnover is
$\approx2.4$ (two-sided) per rebalance, so at 5 bps one-way only the
event- and propagation-based methods keep nonnegative average net Sharpe
ratios (community-aware: $0.10$), and at 10 bps every variant is
unprofitable at this rebalancing frequency. We report this deliberately
rather than tuning it away: it locates the strategy's capacity frontier
and motivates the implementation choices discussed in
Section~\ref{sec:application} (slower rebalancing matched to the
propagation horizon, netting against other signals, and liquid
universes). The Sharpe dispersion is itself a finding: three-year
backtests of modest-IC signals are extremely noisy, and in live research
several non-overlapping validation periods are required before
attributing Sharpe differences of this size to method rather than luck.

\begin{table}[t!]
\centering
\caption{\textbf{Quintile long--short portfolios.} Equal-weight
top-minus-bottom quintile portfolios, rebalanced every 10 trading days;
one-way proportional costs of 5 and 10 bps applied to two-sided turnover
(measured target-to-target at each rebalance). Entries are means across
20 replications (cross-replication standard deviation in parentheses for
Sharpe ratios). Max DD is the gross maximum drawdown including the
inception peak.}
\label{tab:portfolio}
\medskip
\small
\begin{tabular}{lccccccc}
\toprule
Method & Ann.\ ret.\ & Ann.\ vol.\ & Sharpe & Sharpe & Sharpe & Max DD & Turnover \\
 & gross (\%) & (\%) & gross & net 5 bps & net 10 bps & (\%) & (2-sided) \\
\midrule
Sentiment (level) & 1.9 & 4.2 & 0.45 (0.57) & -0.17 (0.57) & -0.79 & 5.1 & 2.07 \\
Direct LLM event & 3.2 & 4.2 & 0.75 (0.46) & 0.04 (0.49) & -0.67 & 4.6 & 2.37 \\
Static-KG propagation & 3.2 & 4.2 & 0.75 (0.46) & 0.04 (0.48) & -0.67 & 4.6 & 2.36 \\
Dynamic-KG propagation & 3.1 & 4.2 & 0.75 (0.55) & 0.02 (0.56) & -0.70 & 4.4 & 2.38 \\
Community-aware propagation & 3.5 & 4.2 & 0.83 (0.50) & 0.10 (0.51) & -0.62 & 4.2 & 2.39 \\
\bottomrule
\end{tabular}
\end{table}

\subsection{What drives the ranking}
\label{sec:sim-why}

The gap between M2 and M4--M5 is the value of cross-entity propagation on
a \emph{current} graph: neighbours' events contain forecastable diffusion
that the direct signal ignores. The M3 result isolates staleness: with
the edge process burned in to its stationary density (so the static
benchmark enjoys no mechanical density advantage), the propagated signal
on a graph frozen at the start of the sample is simply not priced
(median $t=1.1$)---its portfolio behaves like the direct signal's---
because memberships migrate and the ecosystem splits while the frozen
graph cannot follow. The gap between M4 and M5 is the value of the
community gate: down-weighting cross-community edges both matches the
true diffusion asymmetry ($c_{\mathrm{out}}/c_{\mathrm{in}}=0.25$,
versus the a priori $\lout/\lin=0.3$ assumed by M5) and filters
extraction noise, since spurious edges are disproportionately
cross-community. In this calibration the community gate adds about
$0.08$ to the average gross Sharpe ratio and raises the median $\beta_2$
$t$-statistic from $3.4$ to $3.7$ relative to uniform dynamic
propagation; its value would grow with the true asymmetry
$c_{\mathrm{in}}/c_{\mathrm{out}}$, with the spurious-edge rate, and with
the number and separation of communities.
Section~\ref{sec:sim-robust} shows these conclusions are stable across
the operator and detector choices.

\subsection{Robustness of the community gate}
\label{sec:sim-robust}

The community-aware operator involves three researcher choices: the gate
$\lout$, the propagation depth and decay $(K,\gamma)$, and the community
detector. Table~\ref{tab:robust} varies each around the baseline
(10 replications; the composite signal is re-standardized per variant, so
IC levels are comparable within the table). Three conclusions. First, the
gate matters in the hypothesized direction but is \emph{flat near the
optimum}: any $\lout\in[0,0.5]$ performs essentially identically (median
$t\approx3.6$--$3.7$), all dominating the uniform $\lout=1$ operator
(median $t=3.35$)---so the a priori choice of $0.3$ is not doing hidden
work, and a practitioner needs only to down-weight cross-community edges,
not to tune the gate finely. Second, results are insensitive to the
operator's depth and decay ($K\in\{1,2,3\}$, $\gamma\in\{0.2,0.35,0.5\}$
all within $0.1$ of the baseline median $t$), consistent with the one-hop
structure of the true diffusion. Third, the detector is not the binding
choice: greedy modularity maximization performs on par with Louvain, and
even label propagation---a much cruder detector---loses only $\approx0.2$
in median $t$, because the gate only requires the detected partition to
correlate with the true one, not to match it exactly.

\begin{table}[t!]
\centering
\caption{\textbf{Robustness of community-aware propagation.} Each row
re-runs the stylized study (10 replications) varying one ingredient of
the M5 operator around the baseline ($\lout=0.3$, $K=2$, $\gamma=0.35$,
Louvain): mean $h{=}5$ rank IC of the composite signal
(cross-replication s.d.\ in parentheses), the cross-replication median
Fama--MacBeth $t$-statistic on the propagated component, and the fraction
of replications rejecting $H_0:\beta_2=0$ at the $5\%$ level.}
\label{tab:robust}
\medskip
\small
\begin{tabular}{lccc}
\toprule
Variant & Mean IC ($h{=}5$) & Median $t$(Prop.) & Rej.\ rate \\
\midrule
$\lout=0$ (within only) & 0.0184 (0.0039) & 3.71 & 80\% \\
$\lout=0.15$ & 0.0184 (0.0038) & 3.67 & 80\% \\
$\lout=0.3$ (baseline) & 0.0184 (0.0038) & 3.63 & 80\% \\
$\lout=0.5$ & 0.0184 (0.0037) & 3.59 & 80\% \\
$\lout=1$ (uniform, $=$M4) & 0.0180 (0.0036) & 3.35 & 80\% \\
$K=1$ & 0.0184 (0.0035) & 3.54 & 90\% \\
$K=3$ & 0.0185 (0.0037) & 3.67 & 80\% \\
$\gamma=0.2$ & 0.0184 (0.0036) & 3.61 & 90\% \\
$\gamma=0.5$ & 0.0184 (0.0039) & 3.62 & 80\% \\
Greedy modularity detector & 0.0187 (0.0035) & 3.73 & 90\% \\
Label-propagation detector & 0.0183 (0.0035) & 3.41 & 80\% \\
\bottomrule
\end{tabular}
\end{table}

\section{Toward a Russell 1000 Application: Design and a Calibrated
Evaluation}
\label{sec:application}

This section takes the framework toward real markets in two steps. Sections
\ref{sec:app-data}--\ref{sec:app-strategy} give the implementation
blueprint---data, prompts, graph maintenance, and leakage controls---for a
live Russell 1000 study. Section~\ref{sec:app-r1k} then re-runs the entire
Monte Carlo machinery under a \emph{Russell-1000-calibrated} configuration:
the cross-section, network sparsity, volatility structure, news-coverage
distribution, and effect sizes are set to match the stylized facts of the
large-cap U.S. universe, so that magnitudes---ICs, $t$-statistics,
turnover, and cost hurdles---are on the scale a live study would face. We
are explicit about what this is and is not: it is a calibrated simulation,
not an analysis of realized market data, and it therefore validates
\emph{feasibility and magnitudes} rather than the existence of the premium;
the live study specified in this section remains the decisive test.

\subsection{Data and universe}
\label{sec:app-data}

The natural first universe is the Russell 1000 (or MSCI World developed
large/mid caps) to keep transaction costs near the levels assumed above.
Text sources, in decreasing order of signal-to-noise: earnings-call
transcripts (scheduled, information-dense, speaker-attributed); SEC
filings, especially 8-K (material events), 10-K/10-Q (relationship
disclosure: customers, suppliers, risk factors), and S-1s for new
entrants; company guidance and investor-day materials; and point-in-time
financial newswire with capture timestamps. Each document must carry its
\emph{public availability} timestamp, and the extraction pipeline must
process documents in strict timestamp order.

\subsection{Event and relationship extraction}

Appendix~\ref{app:prompts} provides full prompt templates. Three design
rules matter most. First, \emph{extract innovations, not tone}: the prompt
supplies the firm's prior guidance, the previous transcript's key claims,
and consensus estimates, and asks the model to score only the
\emph{unexpected} component ($\Delta$State), returning the tuple
\eqref{eq:event-tuple}. Second, \emph{extract relationships with
direction and evidence}: every proposed edge must quote its supporting
span, so that edge confidence can be audited and hallucinated
relationships filtered. Third, \emph{calibrate}: extraction confidence $c$
should be validated against human labels on a frozen sample before being
used as a weight, and the magnitude scale $m$ anchored with few-shot
exemplars per event category to keep it stationary across time and models.

\subsection{Graph maintenance and community detection}

Edges accumulate evidence scores with exponential decay (half-life of one
to two quarters for news-based evidence; slower for filing-based
relationship disclosures) and are refreshed monthly, matching the
simulation cadence. Louvain with fixed seed per refresh, plus optional
inter-slice coupling \citep{mucha2010community} for stability. Diagnostics
that should be monitored in production: number and size distribution of
communities, modularity level, NMI between consecutive partitions
(structural-change alarm), and case-study inspection of splits---e.g.,
whether an AI-hardware ecosystem separates from legacy semis in the months
its economics diverge.

\subsection{Point-in-time and leakage controls}
\label{sec:application-pit}

The LLM's pretraining corpus may contain the future relative to a backtest
date. Controls, in order of strictness: (i) restrict backtests to periods
after the model's training cutoff (cleanest; limits history); (ii) use
models with documented cutoffs and verify on placebo extractions (ask the
model about post-cutoff events it should not know); (iii) prompt-side
guards instructing the model to use only the supplied documents, with
spot-audits of chain-of-thought for anachronisms. Database joins must be
as-of; the universe must be survivorship-free including delistings; and
corporate actions (mergers, spin-offs, ticker changes) must be resolved in
the entity-resolution layer before graph construction, since a stale
entity map silently corrupts edges.

\subsection{From signals to a strategy}
\label{sec:app-strategy}

The simulation's evaluation stack transfers directly: IC decay determines
the natural holding period; Fama--MacBeth with standard controls
(size, value, momentum, reversal, industry) establishes incrementality;
neutralization of the propagated signal to the direct signal and to sector
exposures isolates the cross-entity component; and capacity is assessed
with realistic cost curves. The comparison set should mirror
Section~\ref{sec:sim}: sentiment, direct events, static-graph, dynamic-graph,
and community-aware propagation, plus the PIS factor for
never-mentioned firms, which is both the cleanest identification of
cross-entity content and a capacity-friendly signal (it concentrates in
less-covered names where diffusion is slowest, cf.\
\citealp{hou2007industry}).

\subsection{A Russell-1000-calibrated evaluation}
\label{sec:app-r1k}

\paragraph{Calibration.} We re-run the full pipeline of
Section~\ref{sec:sim} with every scale-relevant parameter reset to the
large-cap U.S. universe (Appendix~\ref{app:params},
Table~\ref{tab:params-r1k}): $N=1{,}000$ firms over three years;
$K_0=18$ initial communities at industry-group scale with the same
mid-sample ecosystem split; a sparser relationship graph
($p_{\mathrm{in}}=0.10$, $p_{\mathrm{out}}=0.004$, giving roughly $6.5$
within- and $4$ cross-community edges per firm, of supply-chain-database
order); market volatility of $1.0\%$/day ($\approx16\%$ annualized) and
idiosyncratic volatility of $1.6\%$/day ($\approx25\%$); an own-firm
three-day event response of $40$ bps per $1\sigma$ event, with per-edge
neighbour diffusion of $8$ bps within and $2$ bps across communities
(smaller than in the stylized design, reflecting a more efficient
large-cap market); and---new to this configuration---\emph{heterogeneous
news coverage}, with firm-level event intensities drawn from
$U(0.015,0.075)$ per day ($\approx1$--$4.7$ events per quarter), mimicking
the concentration of documents in the most-covered names. The LLM
measurement layer (signal noise, $20\%$ missing edges, spurious edges,
weight noise) is unchanged. We run 10 replications; the larger
cross-section makes per-replication estimates correspondingly tighter.

\begin{figure}[t!]
\centering
\includegraphics[width=0.6\textwidth]{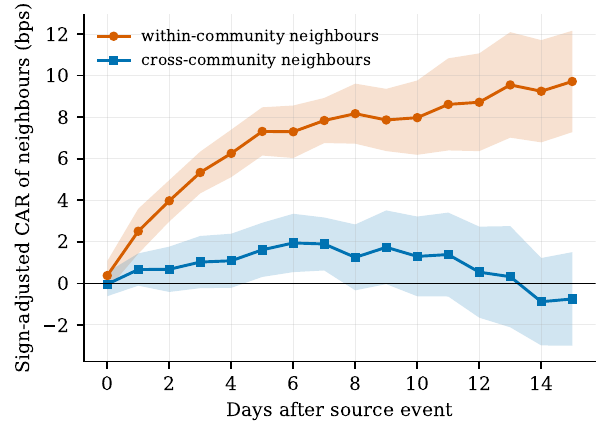}
\caption{\textbf{Cross-entity diffusion at Russell-1000 calibration.}
Sign-adjusted, market-adjusted cumulative average returns of graph
neighbours after source events, split by true community co-membership,
in the Russell-1000-calibrated configuration ($N=1{,}000$; per-edge
responses 8/2 bps). Mean across 10 replications with $95\%$ bands.}
\label{fig:es-r1k}
\end{figure}

\paragraph{Community recovery and mechanism.} The sparser graph is harder:
mean NMI falls to $0.76$ (from $0.86$), and Louvain typically resolves
$15$--$16$ of the $18$ true communities, merging the smallest---a
realistic picture of what industry-group-scale detection on extracted
relationships should achieve. The mechanism remains cleanly identified at
the lower signal level (Figure~\ref{fig:es-r1k}): within-community
neighbours drift to roughly $8$ bps by day 10 after a $|u|\ge0.75$ source
event while cross-community neighbours stay flat, and the feasible pooled
regression \eqref{eq:lambda-reg} gives $\hat\lin=5.4$ bps (s.e.\ $0.6$)
against $\hat\lout=0.8$ bps (s.e.\ $1.5$), a difference $t$-statistic of
$3.6$. The ratio $\hat\lin/c_{\mathrm{in}}\approx0.67$ is again in the
range implied by the attenuation calculus of
Proposition~\ref{prop:attenuation}, with the extra shortfall relative to
the stylized design reflecting the harder detection problem
(misassigned communities blur the within/across split).

\paragraph{Predictive content and incremental pricing.}
Table~\ref{tab:ic-r1k} shows the $h=5$ ICs: uniformly smaller than in the
stylized design, as they should be in a more efficient market---from
$0.0054$ (sentiment) to $0.0096$ (community-aware), with ICIR rising from
$0.17$ to $0.30$---with the same monotone method ordering, and with
community-aware propagation ahead of uniform dynamic propagation
($0.0096$ versus $0.0093$; gross Sharpe $1.16$ versus $1.10$), because at
industry-group scale there are more communities to confuse and more
cross-community edges to discipline. The central test
(Table~\ref{tab:fm-r1k}) survives the move to realistic magnitudes with
honest attrition: the propagated coefficient is significant at the $5\%$
level in $80\%$ of replications for community-aware propagation (median
$t=2.9$) and $60\%$ for uniform dynamic propagation (median $t=2.6$),
against $10\%$ for the static graph (median $t=1.2$); PIS is again
subsumed by edge-level propagation. Three years of daily cross-sections
at true effects of this size is, in other words, \emph{adequate but not
generous} power---a live study should plan for a longer sample, and the
sharper mechanism test \eqref{eq:lambda-reg}, which rejects
$\lin=\lout$ at $t=3.6$ here, should accompany the pricing regression as
a primary outcome.

\begin{table}[t!]
\centering
\caption{\textbf{Rank IC at $h=5$, Russell-1000 calibration.} As
Table~\ref{tab:ic}, for the calibrated configuration; means across 10
replications (cross-replication standard deviations in parentheses).}
\label{tab:ic-r1k}
\medskip
\small
\begin{tabular}{lcccc}
\toprule
Method & Mean IC & IC vol & ICIR & $t$-stat \\
\midrule
Sentiment (level) & 0.0054 (0.0018) & 0.0314 & 0.173 (0.057) & 1.99 \\
Direct LLM event & 0.0074 (0.0022) & 0.0316 & 0.234 (0.071) & 2.70 \\
Static-KG propagation & 0.0083 (0.0024) & 0.0318 & 0.262 (0.076) & 3.02 \\
Dynamic-KG propagation & 0.0093 (0.0028) & 0.0315 & 0.297 (0.089) & 3.42 \\
Community-aware propagation & 0.0096 (0.0029) & 0.0315 & 0.304 (0.092) & 3.51 \\
\bottomrule
\end{tabular}
\end{table}

\begin{table}[t!]
\centering
\caption{\textbf{Fama--MacBeth regressions, Russell-1000 calibration.} As
Table~\ref{tab:fm}, for the calibrated configuration ($N=1{,}000$; 10
replications). The rejection row is the fraction of replications in which
$H_0:\beta_2=0$ is rejected at the $5\%$ level.}
\label{tab:fm-r1k}
\medskip
\small
\begin{tabular}{lcccccc}
\toprule
 & (1) & (2) & (3) & (4) & (5) & (6) \\
\midrule
Sentiment$_{i,t}$ & 0.036$^{***}$ &  &  &  &  &  \\
 & (4.43) &  &  &  &  &  \\[2pt]
Direct$_{i,t}$ &  & 0.053$^{***}$ & 0.052$^{***}$ & 0.050$^{***}$ & 0.049$^{***}$ & 0.049$^{***}$ \\
 &  & (7.19) & (7.11) & (6.69) & (6.63) & (6.66) \\[2pt]
Propagated$_{i,t}$ &  &  & 0.007 & 0.021$^{**}$ & 0.023$^{***}$ & 0.022$^{**}$ \\
 &  &  & (1.18) & (2.55) & (2.91) & (2.41) \\[2pt]
PIS$_{i,t}$ &  &  &  &  &  & 0.003 \\
 &  &  &  &  &  & (0.56) \\[2pt]
Reversal$_{i,t}$ & -0.003 & -0.003 & -0.003 & -0.003 & -0.003 & -0.003 \\
 & (-0.33) & (-0.39) & (-0.38) & (-0.38) & (-0.38) & (-0.38) \\[2pt]
Momentum$_{i,t}$ & -0.001 & -0.000 & -0.000 & -0.000 & -0.000 & -0.000 \\
 & (-0.37) & (-0.30) & (-0.30) & (-0.29) & (-0.28) & (-0.30) \\[2pt]
\midrule
Rej.\ rate $H_0\!:\beta_2{=}0$ & -- & -- & 10\% & 60\% & 80\% & 70\% \\
Avg.\ $R^2$ & 2.6\% & 2.6\% & 2.7\% & 2.7\% & 2.7\% & 2.8\% \\
\bottomrule
\end{tabular}
\end{table}

\paragraph{The cost hurdle, quantified.} Table~\ref{tab:port-r1k} is the
sobering panel and, we would argue, the most useful one for practice.
Gross long--short Sharpe ratios reach $1.16$ for community-aware
propagation, but gross annualized returns are only $1.5$--$2.9\%$ on
$2.5\%$ volatility, while a 10-day-rebalanced quintile portfolio turns
over $\approx2.3$ (two-sided) per rebalance; at 5 bps one-way that is
$\approx2.9\%$ per year in costs, and no variant clears the hurdle as a
standalone strategy---the community-aware method comes closest, at an
essentially breakeven net Sharpe of $-0.03$. The implication is not
that the signal is worthless but that its implementation must respect its
horizon structure: the propagated component decays much more slowly than
the direct component (IC at $h=20$ of $0.0057$ versus $0.0042$ here), so
the natural implementations are (i) longer holding periods with staggered
tranches, (ii) use as an overlay/tilt inside an existing multi-signal
portfolio where its trades net against others, and (iii) concentration in
the low-coverage half of the universe, where diffusion is slowest and PIS
is the only available signal. This is exactly the capacity frontier a
live Russell 1000 study should map, now with quantitative priors for the
power calculation above.

\begin{table}[t!]
\centering
\caption{\textbf{Quintile long--short portfolios, Russell-1000
calibration.} As Table~\ref{tab:portfolio}; means across 10 replications.}
\label{tab:port-r1k}
\medskip
\small
\begin{tabular}{lccccccc}
\toprule
Method & Ann.\ ret.\ & Ann.\ vol.\ & Sharpe & Sharpe & Sharpe & Max DD & Turnover \\
 & gross (\%) & (\%) & gross & net 5 bps & net 10 bps & (\%) & (2-sided) \\
\midrule
Sentiment (level) & 1.5 & 2.5 & 0.61 (0.50) & -0.36 (0.49) & -1.34 & 3.1 & 1.94 \\
Direct LLM event & 2.4 & 2.5 & 0.93 (0.69) & -0.18 (0.71) & -1.29 & 2.8 & 2.22 \\
Static-KG propagation & 2.4 & 2.5 & 0.95 (0.64) & -0.19 (0.64) & -1.32 & 2.8 & 2.24 \\
Dynamic-KG propagation & 2.8 & 2.5 & 1.10 (0.79) & -0.05 (0.79) & -1.20 & 2.8 & 2.29 \\
Community-aware propagation & 2.9 & 2.5 & 1.16 (0.79) & -0.03 (0.80) & -1.22 & 2.6 & 2.33 \\
\bottomrule
\end{tabular}
\end{table}

\section{Discussion and Limitations}
\label{sec:discussion}

\paragraph{What the simulation does and does not establish.} It
establishes that the pipeline is statistically sound: under a known
diffusion mechanism with realistic measurement noise, the feasible
estimator recovers communities, detects structural change, orders the
methods correctly, and prices the propagated signal incrementally. It does
not establish live alpha magnitudes: those depend on extraction quality
(the attenuation in Proposition~\ref{prop:attenuation} applies to both the
signal and the graph), on how much of the diffusion premium remains
unarbitraged, and on crowding.

\paragraph{Model risk in the measurement engine.} LLM extraction errors
are unlikely to be i.i.d.: they correlate with document style, entity
salience, and period-specific narratives, which can induce structured
(not merely attenuating) biases. Ensemble extraction across models,
evidence-span audits, and confidence calibration mitigate but do not
eliminate this. Relatedly, the event taxonomy itself is a modelling choice;
a taxonomy misaligned with how information actually moves would truncate
the signal.

\paragraph{Community model risk.} Modularity maximization has known
pathologies (resolution limit; degenerate near-optimal partitions,
\citealp{fortunato2010community}), and communities are assumed to gate
diffusion symmetrically. The block-matrix generalization
$\phi(k,k')$ and directed, typed propagation are the natural refinements,
at the cost of more parameters to identify.

\paragraph{Endogeneity and reflexivity.} If many investors adopt
graph-propagation signals, diffusion horizons shorten and the measured
$\lin$ shrinks---the strategy consumes its own premium, as the lead--lag
literature suggests has already partially occurred for disclosed
supply-chain links \citep{cohen2008economic}. The dynamic graph is also
partially endogenous to returns if market data enters edge weights; we
recommend text-dominant weighting precisely to preserve exogeneity.

\paragraph{Econometric caveats.} Overlapping returns are handled by
Newey--West, but cross-sectional dependence from the graph itself
(neighbours share shocks by construction) means Fama--MacBeth standard
errors are conservative only under correct specification; block bootstrap
by community is a robustness path. Multiple-testing discipline across the
five-method horse race is addressed here by pre-registration of the
ordering hypothesis and by the 20-replication design.

\section{Conclusion}
\label{sec:conclusion}

This paper develops and validates a framework in which large language
models serve as measurement engines for firm-level economic state
innovations and for the dynamic network through which those innovations
travel. Three design choices distinguish the framework from
entity-level financial NLP: events are measured as innovations
($\Delta$State), the relationship graph is dynamic and text-derived, and
propagation is gated by dynamically detected economic communities with
$\lin>\lout$. In a controlled economy with evolving community structure,
the feasible pipeline detects an emerging ecosystem as the network
rewires and the
community-aware propagated signal is priced incrementally to the direct
signal, with the five-method ordering
(sentiment $<$ direct $<$ static $<$ dynamic $<$ community-aware)
monotone in rank IC across replications and community-aware propagation
attaining the highest long--short Sharpe ratio. At Russell-1000
calibration---a larger, sparser, more efficient configuration with
heterogeneous news coverage---the ordering survives at realistic
magnitudes, the incremental-pricing test rejects in $80\%$ of
replications with the mechanism test sharper still,
and the analysis delivers concrete implementation guidance: the signal's
value lies at longer horizons, in overlays, and in the low-coverage names
where diffusion is slowest. The framework
formalizes a broader hypothesis about modern text-driven markets: alpha
accrues not only to discovering information first, but to modelling how
information propagates across latent economic communities before prices
fully absorb it.

\bibliographystyle{plainnat}
\bibliography{references}

\appendix

\section{Propagation Operator: Properties and Proofs}
\label{app:operator}

\paragraph{Convergence and resolvent form.} Let $P_t$ be any nonnegative
normalization of $\widehat W_t$ with spectral radius $\rho(P_t)$. If
$\gamma\rho(P_t)<1$, the infinite-horizon propagation
$\sum_{k\ge1}\gamma^kP_t^k s_t=\big[(I-\gamma P_t)^{-1}-I\big]s_t$
converges, and the $K$-truncated operator \eqref{eq:baseline-prop} incurs
geometric truncation error
$\|\sum_{k>K}\gamma^kP_t^ks_t\|\le
\frac{(\gamma\rho)^{K+1}}{1-\gamma\rho}\,\kappa\|s_t\|$ for a
diagonalizability constant $\kappa$. With mean-degree scaling
$P_t=\widehat W_t/\bar d_t$ and approximately regular blocks,
$\rho(P_t)\approx1$, so $\gamma=0.35$ gives effective reach of about two
hops---matching the one-hop truth in the simulation DGP with a strictly
positive weight on second-order neighbours as insurance against missing
edges.

\paragraph{Proof of Proposition~\ref{prop:attenuation}.} Write the
regressor $x=\widehat w s=(w+e)(u+\eta)=wu+w\eta+eu+e\eta$ and the
response $y=\lambda wu+\varepsilon$ with $\varepsilon$ independent of all
measurement errors. Under mutual independence and zero means of $(e,\eta)$
with $w\perp u$,
$\mathrm{Cov}(y,x)=\lambda\,\E[w^2]\,\sigma_u^2$ and
$\mathrm{Var}(x)=\E[w^2]\sigma_u^2+\E[w^2]\sigma_\eta^2
+\sigma_e^2\sigma_u^2+\sigma_e^2\sigma_\eta^2
=(\E[w^2]+\sigma_e^2)(\sigma_u^2+\sigma_\eta^2)$
(using $\E[w]=\mu_w$ absorbed into $\E[w^2]$ when $u$ has mean zero;
centring $w$ changes nothing material). Hence
$\hat\lambda\xrightarrow{p}
\lambda\,\frac{\E[w^2]\sigma_u^2}
{(\E[w^2]+\sigma_e^2)(\sigma_u^2+\sigma_\eta^2)}$, which is
\eqref{eq:attenuation}. With the simulation's parameters
($\sigma_u=1$, $\sigma_\eta=0.5$, within-community weights
$w\sim U(0.5,1)$ so $\E[w^2]\approx0.583$, $\sigma_e=0.15$; missing and
spurious edges attenuate further), the predicted factor is
$\approx0.8\times0.96\approx0.77$ before edge-set errors; including the
$h$-day kernel truncation and edge-set errors the feasible
$\hat\lin$ is predicted at roughly $60$--$75\%$ of $c_{\mathrm{in}}$,
consistent with the $\approx73\%$ found in
Section~\ref{sec:sim-mechanism}. $\hfill\square$

\section{Community Detection Details}
\label{app:community}

\paragraph{Louvain.} Starting from singleton communities, phase one moves
each node to the neighbouring community with the largest modularity gain
\begin{equation*}
\Delta Q=\frac{k_{i,\mathrm{in}}}{2m}
-\gamma_{\mathrm{res}}\frac{\Sigma_{\mathrm{tot}}\,k_i}{2m^2},
\end{equation*}
iterating to a local optimum; phase two contracts communities to
super-nodes and repeats. Complexity is near-linear in edges. We use the
NetworkX implementation with \texttt{weight="weight"} and a fixed seed per
refresh; resolution $\gamma_{\mathrm{res}}=1$.

\paragraph{Diagnostics.} Partition quality against ground truth (available
in simulation only) is measured by normalized mutual information,
$\mathrm{NMI}(\mathcal C,\mathcal C^\ast)
=\frac{2I(\mathcal C;\mathcal C^\ast)}
{H(\mathcal C)+H(\mathcal C^\ast)}\in[0,1]$. In production, replace with
temporal self-consistency (NMI between consecutive refreshes) and
modularity tracking; an abrupt drop in self-NMI is the structural-change
alarm that should trigger analyst review of the affected communities.

\paragraph{Alternatives.} Infomap \citep{rosvall2008maps} optimizes
description length of random walks and can be substituted wherever Louvain
appears; degree-corrected SBMs \citep{holland1983stochastic} add a
generative model and principled $K$ selection; multislice modularity
\citep{mucha2010community} couples slices with parameter $\omega$,
trading responsiveness for stability.

\section{LLM Prompt Templates}
\label{app:prompts}

\paragraph{Event extraction (abridged).}
\begin{quote}\small\ttfamily
SYSTEM: You are a financial event-extraction engine. Use ONLY the
documents provided. Do not use knowledge of events after the document
date.\\[4pt]
USER: Company: \{name\} (\{ticker\}). Date: \{date\}.\\
Prior context: \{previous transcript summary; latest guidance; consensus
estimates\}.\\
Document: \{text\}.\\[4pt]
TASK: Identify economic events NEW relative to the prior context. For each
event return JSON:\\
\{category: demand|capex|pricing|inventory|supply\_constraint|guidance|
regulatory|management,\\
\ direction\_magnitude: float in [-3,+3] measuring the UNEXPECTED
component only,\\
\ horizon\_quarters: int, novelty: float in [0,1], confidence: float in
[0,1],\\
\ evidence\_span: exact quote\}.\\
If the document only repeats known information, return an empty list.
\end{quote}

\paragraph{Relationship extraction (abridged).}
\begin{quote}\small\ttfamily
TASK: From the document, extract economic relationships as JSON:\\
\{source\_entity, target\_entity,\\
\ type: supplier|customer|competitor|technology|capex\_exposure|industry,\\
\ direction: source\_to\_target|bidirectional, intensity: float in [0,1],\\
\ confidence: float in [0,1], evidence\_span: exact quote\}.\\
Extract only relationships stated or directly implied by the text; never
infer from general knowledge.
\end{quote}

\paragraph{Anti-leakage guard.} Every call includes the instruction
``\texttt{Treat \{date\} as today; you have no information after this
date}'' and a random subset of calls is audited with placebo questions
about post-date events (a correct pipeline must answer ``unknown'').

\section{Simulation Parameters}
\label{app:params}

\begin{table}[h!]
\centering
\caption{Stylized-design parameters (Section~\ref{sec:sim}; all randomness
from a single seeded generator; the robustness exercise re-runs the full
pipeline for seeds $1,\dots,20$).}
\label{tab:params}
\medskip
\small
\begin{tabular}{lc}
\toprule
Parameter & Value \\
\midrule
Firms $N$ & 300 \\
Trading days $T$ & 750 \\
Initial communities $K_0$ & 6 \\
Ecosystem split day & 375 \\
Within-community edge prob.\ $p_{\mathrm{in}}$ & 0.18 \\
Cross-community edge prob.\ $p_{\mathrm{out}}$ & 0.012 \\
Graph rebuild frequency (days) & 21 \\
Edge survival prob.\ per rebuild & 0.85 \\
Graph burn-in rebuilds (stationarity) & 10 \\
Edge miss rate (extraction) & 0.2 \\
Spurious-edge rate (extraction) & 0.1 \\
Edge weight noise s.d. & 0.15 \\
Event probability (firm--day) & 0.05 \\
Own-firm response $c_{\mathrm{own}}$ (bps) & 50 \\
Within-community per-edge $c_{\mathrm{in}}$ (bps) & 12 \\
Cross-community per-edge $c_{\mathrm{out}}$ (bps) & 3 \\
Neighbour diffusion horizon (days) & 12 \\
Idiosyncratic daily vol. & 0.014 \\
Market factor daily vol. & 0.008 \\
Signal measurement noise s.d. & 0.5 \\
Sentiment level persistence & 0.95 \\
Signal decay half-life (days) & 5 \\
Propagation steps $K$ / decay $\gamma$ & 2 / 0.35 \\
$(\lambda_{\mathrm{in}}, \lambda_{\mathrm{out}})$ used by M5 & (1.0, 0.3) \\
Portfolio rebalance (days) / quantiles & 10 / 5 \\
Transaction costs (bps, one-way) & 5 and 10 \\
Replications (seeds) & 20 \\
\bottomrule
\end{tabular}
\end{table}

\begin{table}[h!]
\centering
\caption{Russell-1000-calibrated parameters
(Section~\ref{sec:app-r1k}; 10 replications). Differences from
Table~\ref{tab:params}: larger and sparser cross-section, industry-group-%
scale communities, large-cap volatility structure, smaller event
responses, and heterogeneous firm-level news coverage.}
\label{tab:params-r1k}
\medskip
\small
\begin{tabular}{lc}
\toprule
Parameter & Value \\
\midrule
Firms $N$ & 1000 \\
Trading days $T$ & 750 \\
Initial communities $K_0$ & 18 \\
Ecosystem split day & 375 \\
Within-community edge prob.\ $p_{\mathrm{in}}$ & 0.1 \\
Cross-community edge prob.\ $p_{\mathrm{out}}$ & 0.004 \\
Graph rebuild frequency (days) & 21 \\
Edge survival prob.\ per rebuild & 0.85 \\
Graph burn-in rebuilds (stationarity) & 10 \\
Edge miss rate (extraction) & 0.2 \\
Spurious-edge rate (extraction) & 0.1 \\
Edge weight noise s.d. & 0.15 \\
Event probability (firm--day) & $U(0.015, 0.075)$ \\
Own-firm response $c_{\mathrm{own}}$ (bps) & 40 \\
Within-community per-edge $c_{\mathrm{in}}$ (bps) & 8 \\
Cross-community per-edge $c_{\mathrm{out}}$ (bps) & 2 \\
Neighbour diffusion horizon (days) & 12 \\
Idiosyncratic daily vol. & 0.016 \\
Market factor daily vol. & 0.01 \\
Signal measurement noise s.d. & 0.5 \\
Sentiment level persistence & 0.95 \\
Signal decay half-life (days) & 5 \\
Propagation steps $K$ / decay $\gamma$ & 2 / 0.35 \\
$(\lambda_{\mathrm{in}}, \lambda_{\mathrm{out}})$ used by M5 & (1.0, 0.3) \\
Portfolio rebalance (days) / quantiles & 10 / 5 \\
Transaction costs (bps, one-way) & 5 and 10 \\
Replications (seeds) & 10 \\
\bottomrule
\end{tabular}
\end{table}

\end{document}